\documentclass[sigconf,screen]{acmart}

\renewcommand\footnotetextcopyrightpermission[1]{} 
\setcopyright{none}
\copyrightyear{2026}
\acmYear{2026}
\acmConference[Tools for Thought at CHI '26]{Tools for Thought at CHI 2026}{April 13--17, 2026}{Barcelona, Spain}

\begin{document}

\title{Guided Sensemaking: Agents in Collaborative Deliberation}

\author{Aaditya Bhatia}
\email{aaditya.bhatia@westpoint.edu}
\orcid{0000-0001-9953-8684}
\affiliation{%
  \institution{United States Military Academy}
  \city{West Point}
  \state{New York}
  \country{USA}
}

\author{Navdeep Kaur Bhatia}
\email{nav@bhatia.dev}
\orcid{0009-0004-5860-4453}
\affiliation{%
  \institution{Independent Researcher}
  \city{West Point}
  \state{New York}
  \country{USA}
}

\author{Marc-Antoine Parent}
\email{maparent@conversence.com}
\orcid{0000-0003-4159-7678}
\affiliation{%
  \institution{Conversence}
  \city{Montréal}
  \state{Québec}
  \country{Canada}
}

\author{Jack Park}
\email{jackpark@topicquests.org}
\orcid{0000-0002-4356-4928}
\affiliation{%
  \institution{TopicQuests}
  \city{Temecula}
  \state{California}
  \country{USA}}


\begin{abstract}
Generative AI systems are aggressively reshaping how students engage with information and perform cognitive work; convenience-oriented use has the potential to displace effortful reasoning, reflection, and learning, especially for those who lack domain expertise and effective human-AI interaction strategies. 
Current AI tools are heavily focused on chat-style interfaces geared towards answer generation and efficiency in a linear and fragmented stream of text, offering limited support for structured reflection, argument construction, and sensemaking in collaborative contexts.
We introduce Guided Sensemaking, an AI-augmented multiagent discourse platform that facilitates composition of well-thought-out ideas around a central question, provides scaffolding for critical thinking, and enables visualization of argumentative structure to support critical thinking and collaborative deliberation.
The system uses several interactive agents to provide context-sensitive questioning prompts and a scaffolding for thought that exposes thematic clusters, agreements, and points of contention without collapsing diverse perspectives.
This paper proposes a conceptual design and interaction paradigm that positions generative AI not as a shortcut to answers but as a research partner that externalizes reasoning, preserves user agency, and fosters structured, traceable sensemaking in educational and civic contexts.
\end{abstract}



\received{12 February 2026}
\received[Accepted]{26 February 2026}
\received[Revised]{26 March 2026}
\received[Presented]{16 April 2026}

\maketitle

\section{Introduction}

Current AI tools primarily focus on generating answers and improving efficiency, but provide limited support for structured reflection and developing reasoning.
Effective sensemaking and learning requires participants to articulate and organize complex ideas. Free form discussions often yield fragmented opinions and hidden assumptions.
This creates a design gap for tools that leverage AI not as a cognitive shortcut, but as an active facilitator of critical engagement that prompts users to refine their reasoning while preserving human agency and learning outcomes.

We present Guided Sensemaking, an AI-augmented interactive system that combines a text composition interface with a visual feedback to enable critical reflection.
A facilitator introduces an overarching topic in the spirit of deliberative platforms like Polis \cite{small_opportunities_2023} and invites participants to respond in long-form text.
As the participants compose their answer, several AI agents engage with them to ask exploratory questions to deepen reflection and visually depict the structure of their entered text.
Combining natural-language interaction with formal issue mapping allows users to answer a top-level question by making claims, posting evidence-backed arguments, or following up with another issue or question.
Users can iteratively ideate before submitting their prose-format text to the ongoing discourse. 
Once submitted, another agent incorporates their contribution into a collective graph that aggregates and connects all users' claims and arguments in the conversation.
The final outcome is a navigable discourse graph \cite{chan2024stepsinfrastructurescholarlysynthesis} representing a well-reasoned group debate supported explicitly by human-generated content.

Our key contributions include a conceptual design and an interaction paradigm that emphasizes human-AI collaboration in structured discourse.
This includes a reasoning scaffold through Socratic questioning \cite{paul_thinkers_2019} and automated development of personal and collaborative discourse graphs.

We illustrate how users engage with the tool and how the discourse graph evolves, thereby presenting a novel way that AI agents can scaffold deliberative thought and boost critical thinking skills towards collective sensemaking.

\section{Related Works}

\subsection{Argument Mining}

Our Reflector agent relies on automatically detecting and extracting claims, premises, support, and attack relations in text, represented by a discourse graph \cite{chan2024stepsinfrastructurescholarlysynthesis}. Compared to a flat text chat, the discourse graph makes logical relationships explicit and preserves the thread of reasoning for later analysis \cite{huang_dagn_2021}. Recent HCI and AI research has begun to combine argument maps with intelligent assistance. Large language models (LLMs) can extract argumentative structure from text in a technique known as argument mining. Anastasiou and De Liddo demonstrated this technique to parse meeting transcripts and automatically generate simplified Issue-Based Information System (IBIS) argument maps \cite{anastasiou_hybrid_2024} that were then applied towards a discussion platform, BCause \cite{anastasiou_shallow_2026}. Bhatia and Sukthankar applied LLMs to policy discourse to turn a large trove of unstructured comments from the Polis platform into an IBIS-style argument tree \cite{bhatia_using_2025} through topic modeling, clustering, and position generation \cite{bhatia_advancing_2024}. They further discussed the suitability of LLMs in moderating large online conversations at scale by assessing relevance to the ongoing discourse \cite{bhatia_moderating_2024}.

\subsection{AI Facilitation and Collaborative Discourse}

Beyond static extraction, structured discourse graphs are highly valuable in active learning and collaboration contexts. Ito et al.\ developed an agent-based conversational system based on a chat interface that maintains the state of the conversation in an IBIS tree and injects facilitation messages, leading to more comprehensive discussion and increased user satisfaction \cite{ito_agent_2022}. Hadfi and Ito further applied the technique to social media to boost deliberation \cite{hadfi_augmented_2022}. In another line of work, Nomura et al.\ created a multiagent chat-based system for collaborative brainstorming and found that humans brainstorming with IBIS-based AI agents produced more ideas per person and reported less hesitation to contribute, compared to human-only brainstorming \cite{nomura_towards_2024}. These results suggest that AI partners backed by structured discourse can actively amplify group creativity and engagement.

\subsection{Impact on Critical Thinking}

Prior research on argument mapping has shown that visual maps help students grasp the structure of arguments and significantly improve comprehension \cite{twardy_argument_2004}. Conversational AI has been extensively used to support and develop these argumentation skills through the concept of cognitive artifacts. Guo et al.\ empirically demonstrated that chatbots designed to debate with undergraduate students led to improved argumentation skills and higher task motivation \cite{guo_effects_2023}.

However, integrating AI into human reasoning requires careful consideration of its impact. Krakauer distinguishes between ``complementary'' cognitive artifacts, which act as coaches that help users internalize mental models and enhance organic intelligence, and ``competitive'' artifacts, which function as automatons that perform tasks without transferring skill \cite{krakauer_will_2016}. He argues that the proliferation of competitive AI tools threatens to diminish human learning and agency by fostering dependency, whereas ideal tools for thought should act as scaffolds that permanently enhance our innate capabilities. Highlighting this risk, Singh et al.\ studied AI-induced cognitive change and noted that instant access to synthesized and complex information can weaken the development of higher-order skills such as analysis, evaluation, and metacognitive reflection when AI substitutes for thinking \cite{singh_protecting_2025}. Supporting this, a recent Microsoft Research study surveyed knowledge workers on their use of generative AI and found that ``higher confidence in GenAI was associated with less critical thinking'' \cite{lee_impact_2025}. To mitigate these risks, researchers like Tankelevitch et al.\ recommend implementing specific strategies in GenAI systems to carefully manage metacognitive load, ensuring workflows promote learning rather than passive reliance \cite{tankelevitch_metacognitive_2024}.

\section{Proposed Mechanism}

\begin{figure}[h]
\centering
\includegraphics[width=\columnwidth]{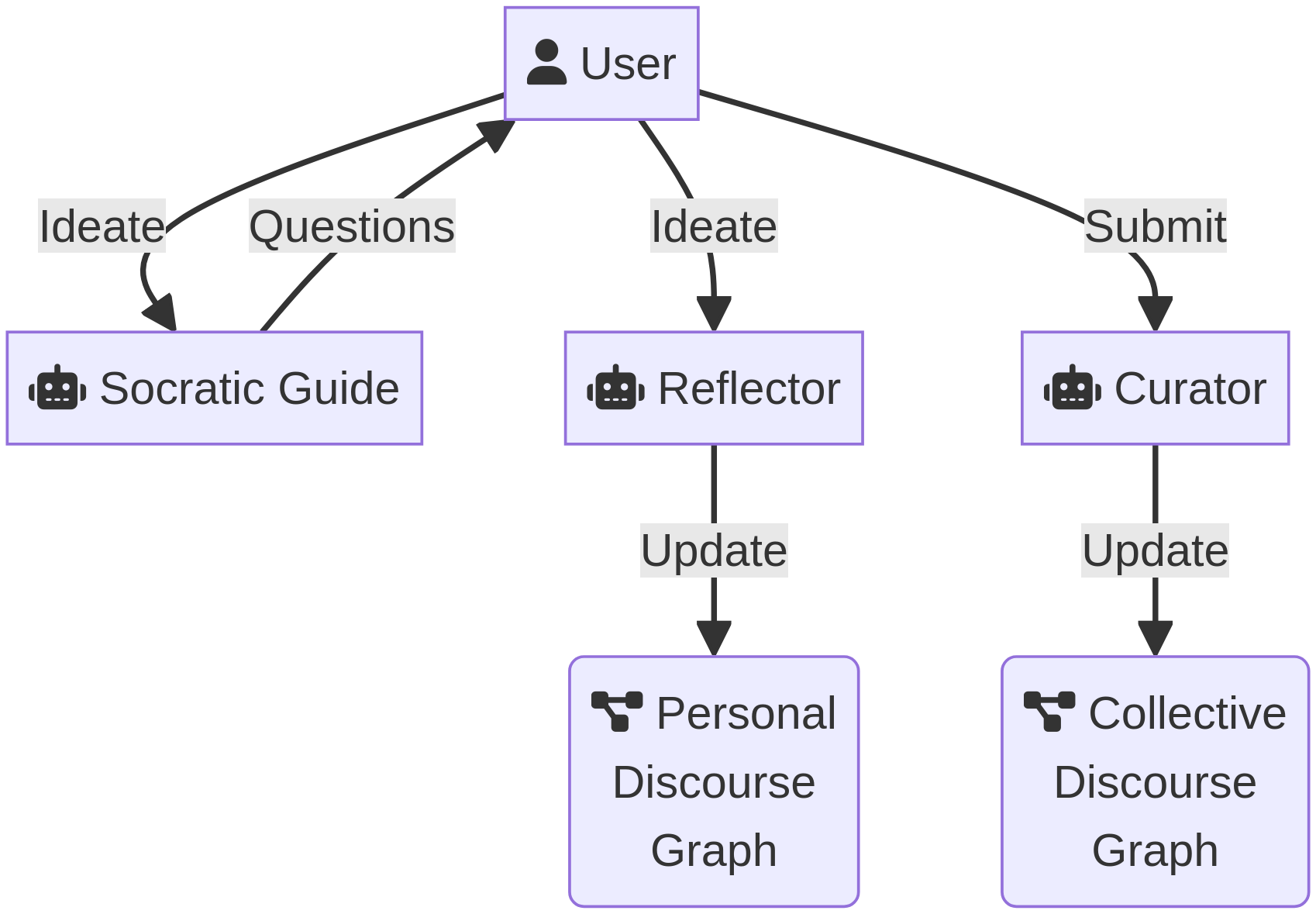}
\caption{User interaction with agents to develop, ideate and submit a contribution.}
\Description[Workflow diagram depicting user interaction]{Workflow diagram depicting the interactions among user, Socratic Guide, Reflector agent, and Curator agent that enable refinement of a user's contributions through critical reflection}
\label{fig:flow}
\end{figure}

\subsection{Interaction Design}

Users interact with Guided Sensemaking through a split interface consisting of an ideation panel and a dynamic visualization as shown in figure \ref{fig:interface}.
The interaction begins when an organizer (e.g., an instructor or facilitator) posts a prompt that frames the overarching topic or problem space.
Participants respond by repeatedly making claims and developing arguments supported by reasoning, examples, and narratives.
Ideation occurs in an iterative manner, where each revision of the write-up causes a Reflector agent to parse the text to extract key claims, assumptions, and supporting elements to update a personal discourse graph. 
This visualization provides a structured, evolving representation of the participant's thinking, including claims, evidence, and inferred relationships among ideas, allowing participants to visually track how their ideas evolve over time. 

In parallel, a Socratic Guide reviews the participant's input and generates targeted prompts intended to surface gaps, ambiguities, or unexamined assumptions. 
The Socratic Guide can also suggest relevant evidence such as scientific publications to aid the research process.
Participants may respond to these Socratic prompts by revising their text, adding clarifications, or incorporating new sources.
However, since this is not a chat-style interface, users cannot directly respond to the Socratic Guide.
This design introduces a deliberate pause between ideating and public posting, encouraging reflection and metacognitive awareness before contributions enter the shared space.

Once satisfied, participants submit their write-up to the collective discourse. 
The Curator agent integrates their contributions into a collaborative graph that represents the structure of the ongoing conversation. 
The agent also performs graph curation operations such as clustering related arguments, identifying overlapping or near-duplicate claims, and merging structurally similar nodes while preserving links to the original posts. 
This results in a high-level visualization of the conversation as it unfolds, supporting sensemaking at both individual and group levels. 
Although the graphical representations and Socratic prompts are generated by agents, the argumentative content itself is authored by participants, preserving user agency while providing structured support for critical reflection.
Users can access their inputs through a dedicated Contributions view, which displays both their own and other participants' submissions, including the final snapshot of associated personal graphs that shows the argumentative structure including claims and supporting evidence. 
This interface illustrates how the collaborative graph evolves as contributions are added. 
Users can explore their inputs within the graph itself by applying filters to identify which claims their contributions support or challenge.
These features provide a comprehensive view of an individual's contributions in relation to those of others.

The core contribution of the system is the use of interactive AI agents that operate continuously alongside participants as ``complementary cognitive artifacts'' \cite{krakauer_will_2016}.
These agents perform several augmentative functions that together support reflective, structured discourse. Figure \ref{fig:flow} depicts the interaction workflow of the system.

\begin{enumerate}

\item \textbf{Reasoning Scaffold}.
The Socratic Guide functions as a lightweight reasoning scaffold during composition.
Rather than supplying content, it poses context-sensitive questions that encourage participants to articulate their claims, consider counterarguments, and assess supporting evidence.
This scaffold is adaptive to the participant's argumentative skills and is designed to prompt deeper reasoning without prescribing conclusions.

\item \textbf{Personal Discourse Graph}. 
The Reflector agent maintains a participant-specific discourse graph that helps visualize their evolving line of thought. 
By translating write-ups into claims, relations, and supporting elements, the graph encourages self-monitoring and reflection, enabling participants to identify inconsistencies, gaps, or over-generalizations in their own reasoning.

\item \textbf{Argument Placement and Graph Curation}.
At the collective level, the Curator agent incorporates individual contributions into a collaborative discourse graph.
It extracts argumentative structure, organizes claims by similarity and relation, links nodes, and preserves traceability to original authors.
This function supports group-level discourse by creating an explicit sensemaking structure without collapsing diverse perspectives into a single synthesized narrative.

\end{enumerate}

Most importantly, users never lose agency; they decide whether to act upon the agent prompts or ignore them, and they always have the final say in how their post is worded. The agents' role is explicitly supportive and auxiliary, reinforcing that both the tool and the human share the goal of better reasoning.

\subsection{Intermediary Outcomes}

As participants interact with the system, several intermediary outcomes emerge prior to the conclusion of the AI agent feedback loop.
At the individual level, participants develop increasingly explicit and well-structured arguments as iterative critique prompts encourage clarification of claims, articulation of assumptions, and consideration of counter-positions.
The personal discourse graph serves as a cognitive artifact that supports reflection-in-action, allowing participants to observe how their reasoning evolves across revisions.

At the group level, the collaborative discourse graph progressively stabilizes into a structured representation of the problem space.
Clusters of related arguments and points of contention become visible, making areas of convergence, divergence, and uncertainty apparent to participants and facilitators.
These intermediary structures enable participants to situate their own contributions relative to others and to identify opportunities for engagement, refinement, or synthesis.

For facilitators, intermediary outcomes include increased visibility into the evolving discourse.
The personal graphs present an opportunity for the classroom instructors to observe and assess student learning outcomes based on the evolution of their thought and ideas.
The collaborative discourse graph provides a diagnostic view of participation patterns, dominant themes, and underexplored perspectives, enabling timely intervention or facilitation when needed.

\subsection{Final Outcomes}

Together, the users produce a collaborative discourse graph with minimal redundancy, clearly depicted argumentative structure, and well-justified contributions.
Rather than producing an overall summary, the Curator agent yields a structured map of claims, arguments, evidence, and relationships that reflects the diversity of participant perspectives and allows the users and facilitators to conceptualize and process the entire problem space.
Each user contributes to this graph in a meaningful, measurable way as all agent-generated graph nodes are derived from the text composed by the users. 

The process ultimately produces a coherent and traceable representation of collective sensemaking around the initial topic.
At the individual level, participants practice developing refined arguments and develop a visual record of their reasoning trajectory, supporting post hoc reflection and learning.
At the collective level, the shared discourse graph functions as a durable boundary object that captures how understanding emerged through interaction and the AI agents are likely perceived as facilitators. 
Through this visualization, users gain a shared understanding of how ideas relate and evolve.
Together, these outcomes position Guided Sensemaking as a tool for reflective discourse by making both individual reasoning and group-level discourse explicit.

\section{Use Case: Classroom Deliberation}

\begin{figure}[t]
\centering
\includegraphics[width=\columnwidth]{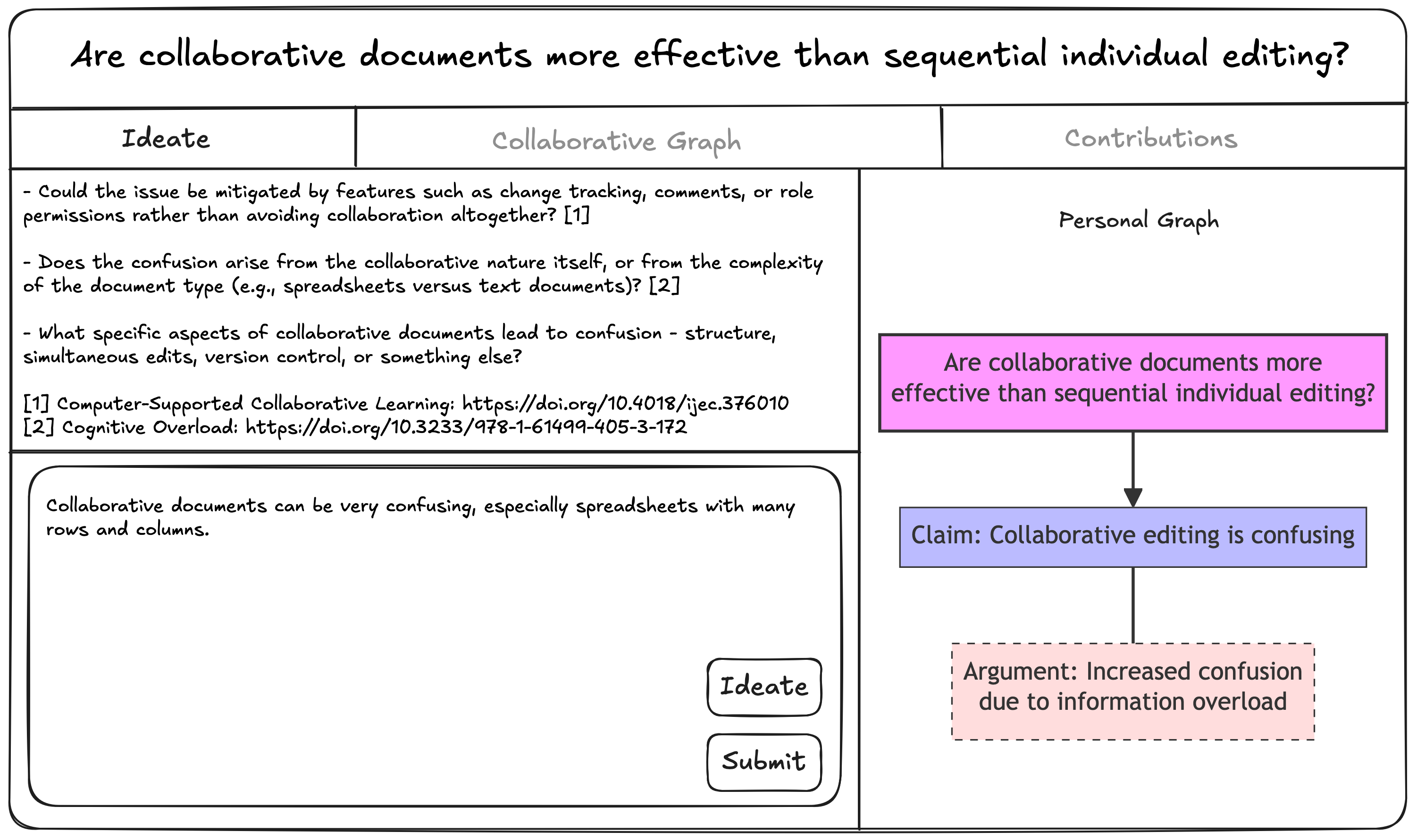}
\caption{User interface for text composition, Socratic guidance, and a personal discourse graph.}
\Description[User interface wireframe.]{User interface wireframe that shows a split interface with a text composition box on the bottom left, Socratic questions immediately above it, and a collaborative discourse graph displayed on the right side.}
\label{fig:interface}
\end{figure}






In a classroom scenario, an instructor posts the question ``Are collaborative documents more effective than sequential individual editing?'' The prompt is framed to invite evaluation rather than factual recall, encouraging students to articulate claims, criteria, arguments, and supporting reasoning.
Students begin brainstorming by going to the Guided Sensemaking platform.
Figure \ref{fig:interface} shows the user interface presented to the students.

\textbf{Step 1.} A student shares their initial thoughts by typing ``Collaborative documents can be very confusing, especially spreadsheets with many rows and columns.'' and clicking on the \textit{Ideate} button.
The Socratic Guide presents the student with the following questions: 

\begin{enumerate}
    \item Could the issue be mitigated by features such as change tracking, comments, or role permissions rather than avoiding collaboration altogether? (agent includes a reference)

    \item Does the confusion arise from the collaborative nature itself, or from the complexity of the document type (e.g., spreadsheets versus text documents)? (agent includes a reference)

    \item What specific aspects of collaborative documents lead to confusion - structure, simultaneous edits, version control, or something else?
\end{enumerate}

\textbf{Step 2.} The Reflector agent extracts claims, evidence, criteria, and arguments to update the personal discourse graph, which helps visualize the evolving ideas. In the given example, the Reflector agent adds the claim ``Collaborative editing is confusing'' to the personal graph and shows the argument ``Increased confusion due to information overload'' as a justification for it. The argument itself remains unsupported.

\textbf{Step 3.} The student revises their write-up, adds clarifications, reviews sources, and cites a study that compares various collaborative systems.
The student may edit claims or arguments and continue to watch their personal discourse graph evolve until they are satisfied with the coherence and strength of their claim. While doing so, the student is able to revisit the history of their personal graph.

\begin{figure}[t]
\centering
\includegraphics[width=\columnwidth]{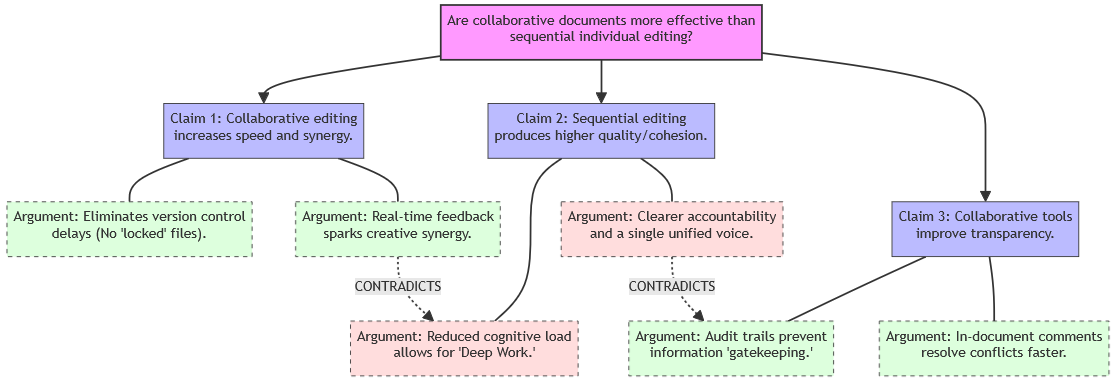}
\caption{Example collective discourse graph depicting a partial problem space around a given issue.}
\Description[Discourse Graph.]{Discourse Graph that shows an issue with two claims, each with several arguments under them.}
\label{fig:graph}
\end{figure}

\textbf{Step 4.} Once ready, the student submits the finalized version of their composed text to the collective discussion.
The Curator agent analyzes the user input to extract the participant's claim, evaluative criteria, and core arguments, and uses these elements to place the contribution into the collaborative discourse graph, associating it with relevant themes and related arguments from other students.
As multiple students submit their thoughts and ideas, the collaborative discourse graph evolves in real time.
The Curator agent clusters overlapping claims that are linked or merged and points of divergence become visible.
Students explore the collaborative graph to see how the discourse evolved, while instructors use it to identify dominant criteria, recurring assumptions, and underrepresented perspectives.
Students apply various filters to the collaborative graph to identify their own contributions in the big picture.
A dedicated \textit{Contributions} view provides a way to explore each contributions in greater depth.
Figure \ref{fig:graph} shows an example of a collective discourse graph.

This use case supports a shift from assumption-driven user responses to evaluative reasoning.
By externalizing individual reasoning during drafting and making collective structure visible after submission, the system encourages students to reflect before contributing and engage more thoughtfully with alternative viewpoints.
The result is a classroom discussion grounded in explicit criteria, traceable claims, and reflective participation rather than rapid opinion exchange.

\section{Conclusion}

Through a conceptual design and an interaction paradigm, Guided Sensemaking demonstrates a practical integration of Socratic questioning techniques and automated discourse graph generation to scaffold deliberative thought.
Agent-assisted metacognitive prompting and visualization helps refine ideas and identify gaps in reasoning.
The system specifically avoids the chat-style fragmented linear conversations and instead the agents act as research partners.
This preserves user agency while making reasoning explicit and navigable at both individual and collective levels.
The resulting personal graph supports reflection-in-action during text composition, while the collaborative graph gradually evolves into a complete representation of the discussion in a given problem space.
Beyond its utility in educational settings, we envision this paradigm serving as a foundation for large-scale civic engagement and collective decision-making for addressing wicked problems.
Future work will include a scoring mechanism to measure the quality of contributions, quality metrics to measure increase in creativity and engagement, and a thorough empirical evaluation of our technique for learning gains, robustness to bias, and scalability in larger deliberations.

\bibliographystyle{ACM-Reference-Format}
\bibliography{references}
\end{document}